\documentclass{emulateapj}
\usepackage{epsfig}
\usepackage{amssymb}
\usepackage{graphicx}

\slugcomment{To appear in ApJ}

\shorttitle{SMA Observations of  RX J1633.9-2442}
\shortauthors{Cieza, L. et al.}

\begin{document}

\title{Submillimeter Array Observations of the RX J1633.9-2442 Transition Disk:  evidence for multiple planets in the making}

\author{Lucas A. Cieza\altaffilmark{1,2},
Geoffrey S. Mathews\altaffilmark{1},
Jonathan P. Williams\altaffilmark{1},
Francois C. M\'enard\altaffilmark{3},
Adam L. Kraus\altaffilmark{1,4},
Matthias R. Schreiber\altaffilmark{5},
Gisela A. Romero\altaffilmark{5,6},
Mariana Orellana\altaffilmark{6}, and
Michael J. Ireland\altaffilmark{7}
 }

\altaffiltext{1}{Institute for Astronomy, University of Hawaii at Manoa,  Honolulu, HI 96822} 
\altaffiltext{2}{\emph{Sagan} Fellow, lcieza@ifa,hawaii.edu}
\altaffiltext{3}{CNRS-INSU, Institut de Plan\'etologie et de Astrophysique de Grenoble (IPAG) UMR 5274, 38041 Grenoble Cedex 9, France}
\altaffiltext{4}{\emph{Hubble} Fellow}
\altaffiltext{5}{Departamento de Fisica y Astronomia, Universidad de Valpara\'{\i}so, Valpara\'{\i}so, Chile}
\altaffiltext{6}{Consejo Nacional de Investigaciones Cient\'{\i}ficas y T\'ecnicas (CONICET), Argentina}
\altaffiltext{7}{Department of Physics and Astronomy, Macquarie University, NSW 2109, Australia}

\begin{abstract}

We present continuum high resolution Submillimeter Array  (SMA) observations of the transition disk object  RX\,J1633.9-2442, which
is located in the Ophiuchus molecular cloud and has recently been identified as a likely site of ongoing giant planet formation.  
The observations were taken at 340 GHz (880 $\mu$m) with  the SMA in its most extended configuration, resulting in an angular
 resolution of  0.3$''$ (35 AU at the distance of the target).  We find that the disk 
 is highly inclined (\textit{i}~$\sim$~50~deg) and has an inner cavity $\sim$25 AU in radius,  which is clearly resolved by our observations. 
We simultaneously model the entire optical to millimeter wavelength spectral energy 
distribution (SED) and SMA visibilities of  RX\,J1633.9-2442 in order to constrain the structure of its disk. 
We find that  an empty cavity $\sim$25 AU in radius  is inconsistent with the excess emission observed at 12, 22, and 24 $\mu$m.
Instead, the mid-IR excess can be modeled by either a narrow, optically thick ring at $\sim$10 AU or an optically thin region
extending from $\sim$7 AU to $\sim$25 AU. 
The inner disk (r $\lesssim$ 5 AU) is mostly depleted of small dust grains  as attested by the lack of detectable near-IR excess.
We also present deep Keck aperture masking  observations in the near-IR, which rule out the presence of a companion
up to 500 times fainter than the primary star (in $K$-band) for projected separations in the 5-20 AU range. 
We argue that the complex structure of the  RX\,J1633.9-2442 disk is best explained by \emph{multiple} planets embedded within the disk. 
We also suggest that the properties and incidence of objects such as  RX\,J1633.9-2442, T~Cha,  
and LkCa 15 (and those of the companions recently identified to these two latter objects)  are most consistent with the runaway 
gas accretion phase of the core accretion model, when giant planets gain their envelopes and suddenly become massive 
enough to open wide gaps in the disk.

\end{abstract}

\section{Introduction}\label{intro}

Observations of nearby star-forming regions show that the vast majority of pre-main-sequence
(PMS) stars are either accreting Classical T Tauri Stars (CTTSs)  with optically thick inner disks  extending 
inward to the dust sublimation radius and showing a narrow range of infrared (IR) colors 
or, more evolved, non-accreting weak-line T Tauri Stars (WTTSs) with bare stellar photospheres. 
The few ``transition objects" that are caught  between the typical CTTSs and WTTSs stages 
present very diverse IR SEDs associated with a wide range of disk structures, 
but they usually have  little or no excess at near-IR wavelength and much larger excesses at longer wavelengths 
(see Williams \& Cieza, 2011 for a recent review). 
Ever since they were discovered by the \emph{Infrared Astronomical Satellite}  (Strom et al. 1989; Wolk \& Walter 1996), the deficit of near-infrared 
excess in transition disks has been interpreted as a diagnostic of inner disk clearing, possibly connected to 
planet formation.  However, in addition to planet formation, at least three other disk evolution processes can  
produce  the inner opacity holes that are characteristic of transition objects: grain growth, photoevaporation, 
and dynamical interactions with (sub)stellar companions. 
The four different mechanisms potentially responsible 
for the holes in transition disks might become distinguishable when, in addition to SEDs, disk masses, accretion rates and
multiplicity information are available (Najita et al. 2007). 
In our recent  multi-wavelength study of Ophiuchus transition disks (Cieza et al. 2010, hereafter Paper I),  we have estimated disk 
masses (from  submillimeter photometry), accretion rates (from  optical echelle spectroscopy), and multiplicity 
information (from near-IR Adaptive Optics imaging)  of 26  transition objects in order to  shed some light on the 
origin of their inner opacity holes.
Of these 26 transition disks,  4  were classified as ``planet-forming disks candidates"
based on their SED morphologies, multiplicity, accretion rates, and disk mass measurements (objects \# 11, 21, 31, and
32 in the study).
These 4 objects have SEDs consistent with sharp, dynamically induced inner holes (as opposed to the smooth 
decrease in opacity expected from grain growth), yet our Adaptive Optics imaging showed that they  lack \emph{stellar} 
companions beyond $\sim$10 AU. 
Also, their large disks masses and/or high accretion rates disfavor photoevaporation as the disk clearing mechanism. 
Overall, the properties of these objects are  those expected for protoplanetary disks with embedded giant planets. 
The recent discoveries of what appear to be forming planets embedded within the disks of the transition objects 
T~\,Cha  (Huelamo et al. 2011) and LkCa 15 (Kraus \& Ireland, 2012) 
give credence to this interpretation and encourages detailed studies of similar objects.

While SED modeling can be a powerful tool to estimate the physical properties of circumstellar disks, 
it has severe limitations. It is a well known problem that different combination of parameters can reproduce the photometry 
data  equally well, even when \emph{very} well sampled SEDs are available (e.g., see Cieza et al. 2011 for the T Cha case).   
Fortunately, resolved (sub)millimeter images can break many of  the degeneracies between model
parameters and help constrain disk properties much better than it is possible with the SED alone  
(Andrews et al. 2009; Brown et al. 2009; Isella et al. 2010).

RX\,J1633.9-2442 is the most massive of the 4 ``planet forming disk" candidates identified in Paper I.
Its SED was recently modeled by Orellana et al.  (2012) as  a 17 M$_{JUP}$ disk  with  a 7.9 AU radius cavity.
Here we present  high resolution   (0.3$''$ $\sim$ 35\,AU) SMA  continuum observations at  340 GHz (880 $\mu$m) 
of RX\,J1633.9-2442
and use the MCFOST radiative transfer code (Pinte et al. 2006) to simultaneously model the entire optical 
to millimeter wavelength SED and SMA visibilities in order to constrain the structure of its  disk. 
In \S 2, we present our SMA data and the photometry from the
literature that we use to construct to the full SED, as well as the Keck aperture masking data we have obtained 
to search for  low-mass companions and the Magellan optical spectroscopy data we use to better measure the
spectral type of  the central star.
Our disk model and the degree to which each 
of the disk parameters can be constrained are discussed in \S 3. 
In \S 4,  we analyze our results 
in the broader context of  disk evolution and planet formation models. 
A summary of our main conclusions
is presented in \S 5.

\section{Observations}

\subsection{\emph{SMA} observations and data reduction}

Submillimeter interferometric observations of our target were conducted in service
mode with the SMA (Ho et al. 2004), on Mauna Kea, Hawaii, 
on February 9$^{th}$  and February 22$^{nd}$,  2010. 
The receivers were tuned to a local oscillator frequency of   340 Ghz (880 $\mu$m).
Both the upper and lower sideband data were used, providing a total bandwidth of 4\,GHz.
The observations were obtained with seven of the eight  6-meter antennas in the 
```very extended configuration",  resulting on 21  baselines from  120 to 510 meter 
in length and  an elongated 
synthesized beam, 0.26$''$$\times$0.31$''$ in size (i.e.,  31 AU $\times$ 37 AU).
The zenith opacities during  both nights  were $\tau_{225\,{\rm GHz}}  \sim$ 0.07.
For each target, the observations cycled rapidly between the target and
two gain calibrators, 1625-254 and 1626-298, located at 2.0 and 5.3 degrees from
RX\,J1633.9-2442, respectively. In order to ensure the appropriate calibration 
of short-timescale phase variations, we adopted  integration times of 
5 minutes on target and 3 minutes on each calibrator.
In each of the two nights,  our  target was observed between  hour angles 
of $-$3.2 and $+$3.4, amounting to a combined \emph{integration} 
time of  7.5 hs. 

The raw visibility data were calibrated with the MIR reduction
package\footnote{available at  \tt  http://cfa-www.harvard.edu/$\sim$cqi/mircook.html}.
The passband was flattened using $\sim$60 to 90 min scans of the bright quasar 0854+201
and the solutions for the  antenna-based complex gains  were obtained using
the primary calibrator 1625-254.  These gains, applied to our secondary calibrator, 1626-298,
served as a consistency check for the solutions.
The absolute flux scale was determined through   observations of Vesta
and is estimated to be accurate to 15$\%$.

\subsubsection{Image plane}

The visibilities were  Fourier transformed, deconvolved  with the 
$CLEAN$ algorithm, and restored with the  synthesized beam using 
the standard $MIRIAD$ software package (Sault et al. 1995). 
Each track was first processed independently to check for consistency, 
but both were later combined to increase the signal to noise of the final image. 
The final image  is shown in Figure 1 and has a rms noise of 1.1 mJy/beam.  
An inspection of the image gives  a first-order  approximation to  some
disk properties. 
First, the image shows two clear peaks, implying that the inner hole
has been resolved in one direction (East-West),  but not in the other.
Also, the aspect ratio of the image suggests the disk is  highly inclined 
(i.e., $\gtrsim$ 45 deg from face-on).
Fortunately, the major axes of the disk and the  synthesized beam are almost perpendicular to 
each other,  maximizing the spatial resolution along the disk. 
The disk is   $\sim$1$''$ (120 AU) across,  and the diameter of the inner hole 
seems to be slightly  larger than the 0.3$''$ (40 AU) beam.
The disk diameter should be considered a lower limit as the SMA observations
are insensitive to the low optical depths of the outermost parts
of the disk (see \S~\ref{modeling}). 
Finally, it is clear that the disk is  not located at the exact center of the field. 

\subsubsection{\textit{uv}-plane}

For interferometric observations, the \textit{uv}-plane 
provides more direct means than the image plane 
to derive quantitative constraints on disk parameters. 
A convenient way to encapsulate the information
from all physical scales sampled by the interferometer is
to deproject the visibilities to 0.0 deg inclination and 0.0 deg 
position angle (PA) for the major axis (Andrews et al. 2009; Brown et al. 2009). 
The deprojected  \textit{uv}-distances are given by 
$R=\sqrt{d^2_a+d_b^2}$, where $d_a =  \sqrt{u^2+v^2}\,  sin\,  \phi$ 
and $d_b =  \sqrt{u^2+v^2}\,  cos\,  \phi \,  cos \, i$,
$\phi = arctan(v/u) - PA$,  and $i$ is the inclination (Lay et al. 1997).
A first approximation to the  
offsets  in $ra$ and $dec$ can be  obtained by examining the  image itself, 
but  more accurate numbers can be calculated by assuming  radial 
symmetry and minimizing the imaginary flux components of
the deprojected visibilities (a face-on radially symmetric 
disk at the center of the field should have zero imaginary flux components
at all spatial frequencies). Using this latter approach, we obtained
positional offsets of $0.17''$ and $-0.24''$ in $ra$ and $dec$, respectively. 
We estimate the PA to be 100$\pm$5 deg  West of North from the orientation of the continuum image
and the inclination  to be 50$\pm$5  deg from the aspect ratio of simulated images (see Section~\ref{empty}).
The deprojected visibility profile, shown in Figure 2, has a null characteristic 
of disks with sharp inner holes (i.e., a large drop in the surface density over a small
radial distance). 
The location of this null is set by the spatial frequency associated with 
the inner hole and directly constrains its radius. 
We binned the data using 40 k$\lambda$ bins. This 
bin size is somewhat arbitrary, but it results in 13  visibility values, 
a number similar to that of the SED points. We have verified that the 
shape of the visibility profile and, in particular,  the location of the null
are robust to the choice of bin size.
In \S~\ref{modeling}, we model the  deprojected visibility profile,
together with the SED, to place constraints on several disk 
structure parameters.

\subsection{Spectral Energy Distribution}

We constructed the optical to millimeter wavelength  
SED for RX\,J1633.9-2442 from the following sources. 
The R-band flux comes from the USNO-B1 catalog\footnote{The USNO-B1
catalog reports two R-band values, 14.67 mag and 15.04 mag; we adopt the average.}
 (Monet et al. 2003), and the ground based near-IR fluxes are from the \emph{2MASS}
survey (Skrutskie et al.  2006). 
The 12 and 22 $\mu$m fluxes come from the 
Wide-field Infrared Survey Explorer (WISE)\footnote{available at: http://irsa.ipac.caltech.edu/Missions/wise.html}.
The \emph{Spitzer}  and 1.3 mm 
fluxes are from Paper I, while the 
850 $\mu$m flux is from Nutter et al. (2006).
Unfortunately, no \emph{Spitzer} spectrum is 
available for this source. 
Our target is embedded in the Ophiuchus molecular cloud, and
thus is strongly affected by extinction. 
Taking advantage of the lack of near-IR excess, 
we estimate an extinction A$_{V}$ = 5.6 mag  from the J-K$_S$  color excess,
adopting A$_V$ = 5.88  $\times$ ((J-K$_S$)-(J-K$_S$)$_o$), 
where (J-K$_o$) is the expected color of a dwarf main 
sequence star  (Kenyon $\&$ Hartmann, 1995) of the same spectral 
type as RX\,J1633.9-2442, a K5 star (see \S~\ref{magellan}).
The extinction at other wavelengths were estimated from the extinction relations 
listed in Cieza et al. (2007). Extinction becomes negligible at 24 $\mu$m and 
beyond, so the long wavelength fluxes have not been corrected for it.
The photometry data from the literature
and the adopted uncertainties, are listed in Table 1.

The resulting SED is shown in Figure 3 and is  characterized 
by the complete lack of detectable excess emission at \emph{2MASS} (1.2 to
2.2 $\mu$m)  and \emph{Spitzer}-IRAC (3.6-8.0 $\mu$m) wavelengths. 
Since a very small amount of dust ($<<$ M$_{lunar}$)
is needed to produce detectable near-IR excess, 
the  SED alone  indicates extreme levels of 
dust depletion in the inner disk (r $\lesssim$ 1 AU) of RX\,J1633.9-2442.
The 22/24 $\mu$m fluxes are somewhat below the lower quartile of 
the CTTS population (Furlan et al. 2006), but the 
70 $\mu$m and (sub)millimeter fluxes are typical 
of a massive primordial disk.

\subsection{Keck aperture masking observations and data reduction}\label{masking}

Non-redundant aperture masking (NRM) has been well-established as a
means of achieving the full diffraction limit of a single telescope
(Tuthill et al. 2000,  2006; Kraus et al. 2008). NRM uses a pupil-plane 
mask to block most of the light
from a target, resampling the primary mirror into set of smaller
subapertures that form a sparse interferometric array. NRM allows for
superior calibration of the stellar primary's point spread function
and elimination of speckle noise by the application of interferometric
analysis techniques, specifically the measurement of closure phases.

We observed RX J1633.9-2442 on April 22-23,  2011,  using the
Keck-II 10-m telescope with Laser-Guide Star Adaptive Optics. All
observations were conducted with the facility AO imager, NIRC2, which
has aperture masks installed in the cold filter wheel near the pupil
stop.  We used a 9-hole aperture mask, which
yields 28 independent baseline triangles about which closure phases
are measured. All NRM observations operate in a subarray mode of the
narrow camera, which has a pixel scale of 9.963 mas/pix, and we
conducted our observations using the broadband $K'$ filter. Each
observing sequence consisted of multiple ``visits'' of RX
J1633.9-2442, alternating with observations of independent calibrator
stars. Many of these calibrators were other transitional disk hosts
in Ophiuchus listed in Paper I. 
Each visit  consisted of a sequence of 12 exposures that were each 20s,
and there were 14 and 17  visits  per night. 

The data analysis was identical to that used in previous papers (e.g.,
Ireland et al. 2008; Kraus et al. 2008, 2011), combined with the new
calibration technique described in Kraus \& Ireland (2012). To briefly
summarize, the images were flat-fielded and bad pixels were removed by
interpolating between neighboring pixels. The image was then
multiplied by a super-Gaussian window function of the form
$\exp(-ar^4)$, with $r$ the radius in pixels from the center of the
interferogram. A two-dimensional Fourier transform was then made of
each exposure in a visit, and this Fourier transform was point-sampled
at the positions corresponding to the baseline vectors in the aperture
mask. For each visit we then computed the vector of mean uncalibrated
closure-phases and the standard error of the mean. Finally, we
calibrated the closure-phases for each visit using an optimal linear
combination of the calibrators observed in the same sequence of
visits.

Our analysis found no statistically significant signal in the
calibrated closure phases for RX J1633.9-2442, and hence that it is
single to within the detection limits of the observations. Using the same
procedures as in our previous NRM work mentioned above (i.e., a 
Monte-Carlo method that simulates  random closure phase datasets of a point source 
with closure-phase errors and covariances that match those of 
the real data), we found contrast limits ($\Delta K'$) of  
5.9 mag at 20-40 mas, 6.9 mag at 40-80 mas, and 6.8 mag at $>$80 mas. 
The corresponding mass detection limits, based on the 1 Myr DUSTY models of 
Chabrier et al. (2000) and the  assumed distance of 120 pc, are 6 $M_{Jup}$ at  
2.4-4.8 AU,  and 3.5 $M_{Jup}$ at $\ge$ 4.8 AU. If  any planetary companions 
are brightened by significant accretion luminosity, as seems likely (see \S~4.4), 
then the mass detection limits could be even lower.

\subsection{Magellan optical spectroscopy observations
and data reduction}\label{magellan}

As part of our recent survey of Ophiuchus transition disks (Paper I), we obtained 
high resolution spectra  of RX J1633.9-2442 using the 2.5-m Du Pont telescope in
Las Campanas  Observatory. From these data, we derived a K7 spectral type and 
an accretion rate of $\sim$10$^{-10}$(M$_{\odot}$yr$^{-1}$).
We reobserved   RX J1633.9-2442  with the Magellan Inamori Kyocera Echelle (MIKE)
spectrograph on the 6.5-m Clay telescope, also  at Las Campanas Observatory,
on June 25$^{th}$, 2011 in order to obtain a second-epoch accretion rate. 

We used the red arm of the spectrograph and a 1$''$ slit to  obtain  the
complete optical spectrum between 4900 and 9500  \AA \   
at a  resolution of 22,000. This resolution corresponds to $\sim$0.3 \AA \ 
at the location of the  H$\alpha$  line and to a  velocity dispersion of $\sim$14 km/s.  
We obtained a set of 3 spectra,  with a exposure  time  of 3  minutes each. 
The data were reduced using the standard IRAF packages  
IMRED:CDDRED and  ECHELLE:DOECSLIT. 

Since the final Clay spectrum has a better signal to noise ratio than our previous 
Du Pont observations,  we have revisited the spectral type classification  of RX~J1633.9-2442
by comparing the new data against the ELODIE high-resolution spectral library  
(Prugniel  \& Soubiran, 2001).
Figure 4 (\textbf{left panel}) shows the RX~J1633.9-2442 spectra in the narrow  6197 to 6203 \AA~region,
containing  V~I and Fe I  lines,  whose shapes, depths,  and relative strengths are highly sensitive
to effective temperature (Padgett et al. 1996). 
We find that the K5 template is a much better match to the RX~J1633.9-2442 spectrum 
than the K7 template is and hence adopt this new and slightly revised spectral type for the modeling work. 

Figure 4 (\textbf{right panel}) shows the continuum subtracted velocity profile of the H$\alpha$ line
for our Du Pont and Clay observations. The vertical lines mark a velocity width ($\Delta V$, 
measured at 10$\%$ of the peak value) of  200 km/s, the boundary between accreting and 
non-accreting objects suggested by Jayawardana et al.  (2003). 
The H$\alpha$ line is broader than 200 km/s, asymmetric and variable in both
shape and intensity as expected from magnetospheric accretion. 
We find $\Delta V$  $\sim$300 km/s from our Du Pont observations and $\sim$230 km/s in our  new Clay data.
For accreting objects, $\Delta V$ correlates  with accretion rates derived from 
models of the magnetospheric accretion process. 
The relation given by Natta et al.  (2004), and adopted in Paper I, 
$Log (M_{acc}(M_{\odot}/yr)) = 
-12.89(\pm0.3) + 9.7(\pm0.7)\times10^{-3} \Delta V (km/s)$, 
translate the $\Delta V$ values into accretion rates of 10$^{-10}$ and   10$^{-10.6}$
 M$_{\odot}$yr$^{-1}$, for the Du Pont and Clay data, respectively.

\section{Disk model}\label{modeling}

In order to constrain the structure of the RX J1633.9-2442 disk based on
the observed SMA visibilities and  SED, we use the MCFOST radiative 
transfer code  (Pinte et al. 2006). MCFOST adopts a Monte Carlo approach to 
follow   ``photon packets'' propagated through the disk  
(i.e., a parametrized dust density structure). 
MCFOST outputs synthetic SEDs and monochromatic raytraced images. 
The raytraced images can be used to simulate  synthetic SMA visibilities with the same \textit{u-v} sampling
as the actual observations.  The deprojected synthetic visibilities can then be directly compared against the
real  data, using the same radial binning,  as described in the following section.

\subsection{Fitting procedure}

We follow the model fitting procedure described by Mathews et al. (2012), which uses
the Levenberg-Marquardt  $\chi^{2}$ minimization algorithm to perform an efficient exploration
of the parameter space. 
We start by using MCFOST to create a small grid of models and the  corresponding SEDs and 880 $\mu$m raytraced images.  
The raytraced images are input to  the FT  (Fourier Transform) task  in the  Common Astronomy Software Applications 
(CASA\footnote{available at http://casa.nrao.edu/}) package,
which outputs  visibility data sets with the same \textit{u-v} sampling as the real SMA observations.
Then, for each model, the $\chi^2$ of the model is calculated as the sum of the $\chi^2$ of the SED 
and the $\chi^2$ of the visibility profile. 
We use the IDL routine MPFIT (Markwardt, 2009) to implement the Levenberg-Marquardt  $\chi^{2}$ minimization algorithm 
to calculate the numerical gradients of the $\chi^2$ function and determine the next point in the parameter space to be 
sampled until the algorithm converges to a $\chi^2$ minimum.

To better sample the parameter space and in order to avoid local minima, we carried out the search algorithm 
several times using different starting values. Each of the runs  provide a set of best-fit parameters. The distribution
of the  best-fit values for each parameter can be used to calculate the mean and an associated uncertainty 
(see \S~4.2 in Mathews et al. 2012 for a discussion on estimating the uncertainties). 

\subsection{Disk parameterization}\label{paramet}     

Transition disks are complex systems that have been modeled with a 
wide range of structures.
Some  objects such as CoKu Tau/4  and DM Tau 
have inner holes that seem to be completely depleted of 
IR emitting grains (D'Alessio et al. 2005; Calvet et al. 2005),
while others such as GM Aur have cavities filled with optically thin 
dust (Calvet et al. 2005).  Yet other systems,  like T Cha and LkCa 15 
are best described as having optically thin gaps separating optically 
thick inner and outer disk components (Olofsson et al. 2011; Espaillat et al. 2010).
In what follows, we explore all  3 possibilities for the structure of 
the RX J1633.9-2442 disk.

\subsubsection{Empty cavity}\label{empty}

We begin our modeling by adopting the simplest possible structure for a transition disk: 
a disk with an empty cavity. 
Motivated by physical models of viscous accretion disks (e.g., 
Hartmann et al. 1998, Lynden-Bell \& Pringle 1974)  and  
the discrepancy in disk sizes obtained from continuum and CO
line images (Hughes et al. 2008; Isella et al. 2007; P\'ietu, Guilloteau \& Dutrey, 2005),
we follow  Mathews et al. (2012) and adopt the following description for the surface 
density  profile and  the size of the disk:  
 
$ \Sigma_{dust}(R)=\Sigma_{C} \left({R\over R_c}\right)^{-\gamma}\,
          \exp\left[-\left({R\over R_c}\right)^{2-\gamma}\right]$

where $\Sigma_{C}$ is  the surface density at a characteristic radius, $R_c$, 
and $\gamma$  is the radial dependence of the disk viscosity,
$\nu\propto R^\gamma$.  In this prescription, the surface density  
is proportional to R$^{-\gamma}$ in the inner disk, but it quickly
becomes dominated by the exponential taper at large radii.
$\Sigma_{dust}$ = 0 within the cavity of radius $R_{cav}$.
We set  the outer radius  of the disk, $R_{out}$, to 200 AU.
However, we note that our model is not very sensitive to the exact value
of  $R_{out}$ because the exponential taper of the outer disk implies very 
low surface densities (below the sensitivity of our SMA observations) for
radii $>>$ $R_{C}$. 
The vertical distribution of the dust is given by a Gaussian with a scale height 
$h~=~h_CR_C(R/R_C)^{1+\psi}$, where $h_CR_C$ is the scale height at $R_C$,  and 
$\psi$ is the power law describing the flaring of the disk.
The dust content is described by a differential power-law for the grain size distribution
(d$n$($a$) $\propto$  $a^{-p}$d$a$), between  0.005 $\mu$m and 3900 
$\mu$m.\footnote{the maximum grain size is somewhat arbitrary, but corresponds to 3$\times$
the size of the longest wavelength in  the SED, which approximates
the maximum grain sizes the data are sensitive to (Draine 2006).}
We adopt the porous grains from Mathis $\&$ Whiffen (1989) for the grain composition.
The stellar parameter are those of  the best matched photospheric model (Kurucz 1979, 1993) 
for a  K5 star  (T$_{eff}$=4350 K, log g = 4.0)  at the relatively well established distance  of 
120 pc to the Ophiuchus  molecular cloud  (Loinard et al. 2008). 
We initially created SEDs and visibility profiles assuming an inclination of 60 deg
for the disk, but quickly found that adopting an inclination of 50 deg resulted in synthetic 
images with aspect ratios that are closer to the observed image and hence use this 
latter value for all models. 
Table 2 lists all the parameters that are fixed in our model. 
The 6 free parameters for the ``empty cavity" model are listed in Table 3.
M$_{disk}$ is not a free parameter. It is obtained by integrating the surface
density profile over radius and assuming a gas to dust mass ratio of 100.

For this parameterization,  we ran the search algorithm 5 times. 
The visibility profile and the SED of the overall best-fit model from all 5 runs  
are indicated in Figures 2 and 3, respectively.  
The probability weighted means and uncertainties (calculated as in
Mathews et al. 2012) for each parameter are listed in Table 3 in the ``Empty cavity"  column.
While the model reproduces the visibility profile  very well, 
the observed 22 and 24 $\mu$m fluxes are factors of $\sim$2-3 
(i.e., $\sim$10 to 20$-\sigma$)  higher then predicted by the model. 
Since the 22 and 24 $\mu$m measurements are independent 
(the former is from WISE and the latter is from \emph{Spitzer}), photometric 
problems can be ruled out. 
The ``empty cavity" model does not reproduce the significant 12 $\mu$m excess either. 
The large discrepancy between the observations and the best-fit model in the mid-IR
can be understood considering that the search algorithm was most likely driven
by the $\chi^2$ of the visibility profile, which is very sensitive to the value of 
R$_{cav}$. 
In other words, the visibility profile  very strongly constrains 
the size of the inner cavity to be 22.7$\pm$1.6 AU, but 
such a large cavity is incompatible with the observed 12 and 22/24 $\mu$m
excesses. 
We find that the SED and  SMA data  can \emph{not} be reconciled even adopting 
a  hotter K1-type  central star  (which is clearly ruled out by the optical spectrum, see Figure 4).  
It is thus unavoidable to conclude that the  $\sim$23 AU cavity imaged
at submillimeter wavelengths is not completely depleted of 
mid-IR emitting grains.

\subsubsection{2-component disk} 

To try to reproduce the observed mid-IR excesses,  we partially fill the cavity  by incorporating 
2 additional free parameters that result in a 2-component disk: $\delta_{cav}$ and $R_{in}$. 
 Following Andrews et al. (2011a) and Mathews et al.  (2012), the surface density profile of the 
 disk is modified such that  $\Sigma_{dust,cav}$~=~$\delta_{cav}$~$\Sigma_{dust}$,  
 for $R_{in}$ $<$ R $<$ $R_{cav}$. That is, $\Sigma_{dust}$  no longer  drops 
 to zero at $R_{cav}$, but is sharply reduced to a lower value between $R_{cav}$ and $R_{in}$. 
This sharp reduction in the surface density profile at $R_{cav}$ is meant
to reproduce the inner hole seen at submillimeter wavelengths.   
$\Sigma_{dust}$ remains  zero for R $<$ R$_{in}$.

We ran the search algorithm 5 times for this new parameterization.  
The visibility profile and the SED of the overall  best-fit 2-component
model are indicated in Figures 2 and 3, respectively.  The weighted means and uncertainties 
for each parameter  are listed in Table 3 in the ``2-component disk"  column.
The inner component of this model is characterized by a surface density reduction of 
$\sim$100 with respect to the outer disk and an inner radius of $\sim$7 AU. 
As shown by Figures 2 and 3, the 2-component parametrization allow us 
to simultaneously obtain satisfactory fits for both  the visibility profile 
and the SED. 
However, as the properties of the inner disk component are effectively controlled 
by only 3 SED points (i.e, the 8.0, 12, and 22/24 $\mu$m fluxes), the solution is 
unlikely to be unique.  We next explore an alternative geometry for the
inner disk, a narrow ring. 

\subsubsection{Narrow ring within cavity}

As an alternative of the 2-component model,  we modify the empty cavity model with
two additional parameters, $R_{ring,in}$ and $R_{ring,width}$, describing the location
and  width of a ring within the cavity. The description of the outer disk (i.e., beyond R$_{cav}$) 
remains unchanged.  
With this parametrization, we also ran the search algorithm 5 times. 
The results are shown in Figures 2, 3 and listed in Table 3, together 
with the other 2 earlier  parameterizations.
We find that a narrow ring at $\sim$10 AU fits  the visibility profile 
and the SED almost as well as the  2-component model does
(see also $\chi^2$ of the visibilities and the SED in Table 3).
The small width of the ring (2  AU) can be understood from the fact
that the ring is optically thick and its mid-IR emission is dominated 
by the inner rim facing the star. 

\subsection{Synthesis of Modeling results}~\label{mod-results}

Table 3 shows that the outer disk parameters for  all three model structures 
agree remarkably well. 
All the values are well  within 1-$\sigma$ or 2-$\sigma$, which 
give us confidence in both  our modeling results  and uncertainty estimates.   
Also, because the SMA data is only sensitive to the properties 
of the outer disk, the visibility profiles of the three models match the 
observations comparably well (see Figure~2 and $\chi^2$ of the visibilities in Table 3). 
We find that  the disk is relatively massive ($\sim$15 M$_{JUP}$) 
and  rather flat. The scale height at R$_C$ is given by $R_C \times h_C$, 
corresponding to $\lesssim$ 0.8 AU at a radius of  $\sim$40 AU for all 3 models.  
This flat geometry is in agreement with the result by Orellana et al. (2012),
who found a scale height of 2 AU at 100  AU for RX~\,J1633.9-2442, and is most
likely due to significant grain growth and dust settling. 
Our models, driven by  the (sub)millimeter colors, do favor a grain size distribution
extending beyond the millimeter size scale. 

As for the the inner disk, its  structure is  not well constrained by current  data.
The narrow optically thick ring  and the more extended optically  thin region
are only a subset of possible solutions,  and the radial symmetry  assumed 
is not necessarily  correct  (see \S~\ref{dynamical}).  
However, we can say that the mid-IR emission originates beyond 
$\sim$5 AU, from a dust component that is distinct from the outer disk imaged
at submillimeter wavelengths. Similarly, the extreme depletion of  dust grains 
within a few AU of the star is well established by the lack of detectable excess
at IRAC wavelengths (3.6 to 8.0 $\mu$m). 
As mentioned in the introduction, Orellana et al.  (2012) successfully modeled 
the SED of  RX~\,J1633.9-2442 with a 7.9 AU inner hole, which was virtually empty. 
In other words, the larger submillimeter cavity imaged by the SMA is \emph{not} 
detectable from the SED alone.
The surface density profile of the 3 models we considered are shown in Figure~\ref{sigma}.
By simultaneously modeling the visibility and SED data,  we are able to demonstrate
that the circumstellar environment of RX~\,J1633.9-2442 presents \emph{at least} 
three distinct  radial regions. The innermost  region (r $\lesssim$ 5 AU)   
is depleted of small grains. 
The middle region contains \emph{some} mid-IR emitting grains,  with  an unknown 
configuration. 
The outermost region of the disk starts at $\sim$25 AU, with a sudden increase 
in the surface density.  
This complex structure is certainly intriguing, but is not unique 
to RX~\,J1633.9-2442.  
Similar structures have  already  been proposed to reconcile the 
submillimeter images and SEDs of several other transition
disks, including DM Tau,  RX~J1615.3-3255 (Andrews et al. 2011a),
and  RX~J1604.3-2130 (Mathews et al. 2012).  
In the next section, we discuss the physical processes that could 
potentially  explain our modeling results and the overall properties 
of RX~\,J1633.9-2442. 
Most of our conclusions are also applicable to the 3 other similar 
objects listed above. 

\section{Discussion}

As discussed in \S~\ref{intro},  multiple mechanisms 
have been proposed to explain the inner opacity holes
of transition disks, including photoevaporation, grain growth,
and dynamical clearing.  In what follows, we consider how the 
predictions from models of each  process compare to the
properties of  RX~\,J1633.9-2442.
Toward the end of the section,  we  assess the likelihood
of a connection between  transition disks and each of 
the two leading  theories of giant planet formation, core accretion 
and gravitational instability. 

\subsection{Photoevaporation}

Photoevaporation by the central star is currently believed to play an important 
role on the dissipation of circumstellar disks. 
Photoevaporation can be driven by energetic photons in the FUV (6\,eV $< h\nu <$  13.6\,eV),
EUV (13.6\,eV $< h\nu <$ 0.1\,keV ) and X-ray ($h\nu > 0.1$\,keV) energy range.
Photons in each energy domain operate in different ways.
EUV photons cannot penetrate far into the disk and drive relatively weak 
photovaporation winds  (10$^{-10}$ M$_{\odot}$/yr).
Thus, EUV photoevaporation  only becomes important  
once most of the disk mass has been depleted and the accretion rate drops 
below $\sim$10$^{-10}$  M$_{\odot}$/yr (Alexander et al. 2006a,2006b),
at which point the outer disk is no longer able to resupply the inner
disk with material and the inner disk drains on a viscous
timescale ($\sim 10^5$\,yr). 
Once the inner disk is drained, a hole is formed, the  disk edge is directly exposed 
to the EUV  radiation, and the disk rapidly photoevaporates from the inside out.
The very large disk mass of  RX~\,J1633.9-2442  and the presence of 
accretion are inconsistent with EUV-induced photoevaporation being 
the formation mechanism for its inner hole. 

More recent studies have incorporated   X-ray (Owen et al. 2011; 2012) and/or
FUV irradiation (Gorti, Dullemond \& Hollenbach 2009; Gorti \& Hollenbach 2009)
into photoevaporation  models.  
According to these models,  X-ray and FUV photons can penetrate deeper into 
the disk, and drive much higher photoevaporation rates ($\sim$10$^{-8}$
M$_{\odot}$/yr). As a result, the hole is expected to form earlier in the 
evolution of the disk, while the disk is still relatively massive. Since
the surface density of the inner disk at the time it starts to drain is high,
X-ray/FUV photoevaporation models can in principle explain the 
presence of massive disks with inner holes and moderate accretion
(i.e., they would represent the inner disk draining stage).
However, these models also predict that  accretion onto the star should quickly 
drop as the size of  the inner cavity grows.

In the context of  these models, whether the inner hole of a transition disk can potentially  be explained by X-ray/FUV 
photoevaporation depends on the size of the inner hole, the observed accretion rate, 
the stellar mass, and the X-ray luminosity. 
In particular, the region in the hole size versus accretion rate plane that is
consistent with X-ray/FUV photoevaporation is a very strong function
of stellar mass, M$_{\star}$, (see Fig. 17 in  Owen et al. 2012) because:  1) the location at which
the X-ray heated gas becomes unbound and opens a gap in the disk  is proportional to M$_{\star}$
and 2) the X-ray luminosity is also a strong function of stellar mass (L$_{X-ray}$ 
$\propto$ M$_{\star}^{3/2}$; Preibisch et al.  2005). 
We thus estimate the stellar mass of  RX~\,J1633.9-2442 by comparing its 
temperature and luminosity to the predictions of theoretical evolutionary tracks. 
A T$_{eff}$ of  4350 K is directly derived from the K5 spectral type. 
We estimate a luminosity of 0.70 L$_{\odot}$ applying a bolometric correction (from Hartigan
et al. 1994)  to the extinction corrected J-band magnitude and  adopting a distance of 120 pc.
According to the  models by D$'$Antona $\&$ Mazzitelli (1998) and  Siess et al. (2000)
the temperature and luminosity of RX~\,J1633.9-2442  correspond to those of a 2 Myr old  0.7
M$_{\odot}$ star  and a 6 Myr 1.0 M$_{\odot}$ star, respectively. 
The large discrepancy in stellar age highlights the uncertainty of these models.  
However, with that caveat,  RX~\,J1633.9-2442  can be considered to be a 
$\lesssim$ 1 M$_{\odot}$ star, for which disk  holes  larger than 20 AU around 
accreting objects cannot be explained by photoevaporation.

Also, photoeveporation cannot easily account for the 3-region structure we found in \S 3.
We thus conclude that the inner hole of  RX~\,J1633.9-2442 is unlikely to be due 
to any kind of photoevaporation process. 

\subsection{Grain growth}

Dust opacity,  $\kappa_{\nu}$  (cm$^2$/g),  is  a very  strong  function of particle size. 
As soon as primordial sub-micron dust grains grow into  larger bodies
(r $>>$ $\lambda$), most of the solid mass never interacts with the radiation, 
and  $\kappa_{\nu}$  plunges. 
Observational support for grain growth in disks is robust and comes from at least
two independent lines of evidence: the shapes of the silicate features around 10 and 
20 $\mu$m (Kessler-Silacci; 2006, Olofsson et al. 2010) and the spectral slopes of disks 
at (sub)millimeter wavelengths (Andrews \& Williams, 2005, 2007; Wilner  et al. 2005; Ricci et  al.  2010). 
Grain growth  has been proposed as one of the possible explanations for the opacity holes
of transition disks  because it  might be a strong function of radius  (it is 
expected to be more efficient in the inner regions where the surface density is higher and the 
dynamical timescales  are shorter).

Idealized dust coagulation models, ignoring fragmentation and radial drift, do in fact predict extremely 
efficient grain growth in the inner disk and can produce  SEDs similar to those of RX~\,J1633.9-2442
(Dullemond \& Dominik, 2005). 
However,  dust fragmentation and radial drift result in the efficient replenishment of
micron size grains (Brauer, Dullemond \& Henning 2008; Birnstiel, Ormel \& Dullemond 2011)
and a smooth and \emph{shallow} dependence of   $\kappa_{\nu}$  on disk radius. 
In contrast,  both the SED and SMA visibilities of  RX~\,J1633.9-2442 are consistent with 
a \emph{steep} discontinuity in the optical depth (i.e.,  $\kappa_{\nu}\times\Sigma$) at R$_{cav}$ and R$_{in}$ (or R$_{ring,in}$).
Furthermore,  if  the  optical depth discontinuities are  mainly  due to a reduction in $\kappa_{\nu}$ instead of 
$\Sigma$, this would favor the onset of the magneto-rotational instability (Chiang  \& Murray-Clay, 2007)
and would exacerbate  accretion. 
The very low accretion rate of the  RX~\,J1633.9-2442 disk ($\lesssim$ 10$^{-10}$ M$_{\odot}$yr$^{-1}$; see \S~\ref{magellan})  
contradicts this scenario. 
We thus consider grain growth to be a very unlikely explanation for the inner opacity reductions. 

\subsection{Dynamical clearing}\label{dynamical}

Unlike grain growth,  the dynamical interaction of a (sub)stellar 
companion embedded within the disk can produce a sharp inner 
hole (Artymowicz \& Lubow (1994). 
Ireland $\&$ Kraus (2008)  showed that the famous transition disk CoKu~\,Tau~\,4  is in fact  a near-equal mass binary system,
which naturally explains the hole that had been inferred from its SED. This immediately raised  the question of whether  most
transition disks were close binaries.   
However,  it is now clear that most sharp holes are \emph{not} due to binarity.  The transition objects DM Tau, GM Aur, LkCa 15, UX Tau,
and RY Tau have all been observed with the Keck interferometer (Pott et al. 2010). For these objects, stellar companions with flux ratios
  $<$ 20 can be ruled out down to sub-AU separations. 

Our  Keck aperture masking observations discussed in \S~\ref{masking} rule out the presence of 
even a brown dwarf companion down to a projected separation of 2.4 AU (corresponding to 
a maximum physical separation of 3.8 AU).  
Therefore, a stellar (or brown dwarf)  companion cannot explain neither the
 $\sim$25 AU hole seen in our  SMA image nor  the inner most cavity indicated by the SED. 
Instead,  the 3-region structure of the RX~\,J1633.9-2442 disk discussed in \S~\ref{mod-results} suggests the
presence of  \emph{at least}  two low-mass objects  dynamically sculpting the disk:  an object at $\lesssim~25$~AU 
from the star that creates the  discontinuity in the surface density seen in  the submillimeter image and another 
object at $\lesssim$~7~AU that clears up the inner disk of  material. 
We thus argue that the dynamical interaction of multiple planets embedded within the disk is the most likely
explanation for the overall properties of RX~\,J1633.9-2442.

The need for multiple planets to explain the properties of RX~\,J1633.9-2442 is very strongly supported by
recent hydrodynamical simulations  of giant planets embedded in primordial disks by  Dodson-Robinson 
\& Salyk   (2011)  and  Zhu et al.  (2011) 
showing  that multiple planets are in fact \emph{required} to produce inner holes and 
gaps wide enough to have a noticeable effect in the emerging SED. 
Both studies find that a single giant planet cannot explain wide optically thin gaps and holes. 
Multiple forming planets also help explaining the low accretion rates \emph{onto the star}  (for a given disk mass) of 
many transition objects  (Najita et al. 2007; Espaillat et al. 2012) as each planet accretes a significant fraction
of the material being transported across the disk.

\emph{Kepler} observations  have demonstrated that systems with multiple planets with small semi-major axes
are common (Lissauer et al. 2012).  
The lack of near-IR excess ($\lambda <  $ 8.0 $\mu$m)  in the SED of  RX~\,J1633.9-2442 combined with 
variable and detectable levels of accretion ($\sim$10$^{-10}$--10$^{-10.5}$ M$_{\odot}$yr$^{-1}$) could
be an indication of a densely packed  planetary system resulting in multiple optically thick tidal streams 
that transport a significant amount of material onto the star but cover a small area of the 
inner disk. 
We ran some tests and found that  \emph{axisymmetrically distributed} circumstellar material would produce 
a detectable near-IR excess unless the surface density  of the inner disk (r~$\lesssim$~5~AU) 
is reduced by a factor of 10$^4$ with respect to that of a ``typical"  CTTS disk. 
In the absence of planets,  the accretion rate onto the star should be proportional to the surface density of the 
inner disk. Therefore, if the surface density is reduced by a 10$^4$  factor from typical levels,  one would expect 
an accretion rate of  10$^{-12}$--10$^{-12.5}$  M$_{\odot}$yr$^{-1}$ for RX~\,J1633.9-2442.  Such low rates are 
undetectable and 2 orders of magnitude lower than the observed value. 
Nevertheless, the lack of near-IR excess and the accretion rate could be reconciled if  the inner disk contains
optically thick tidal streams with a geometric filling factor of a few percent, which is in 
agreement with  the results of hydrodynamic simulations (Dodson-Robinson \& Salyk,   2011).
Since these hydrodynamic simulations 
predict very complex  disk structures, the ``2-component" and  ``ring within cavity" models presented herein are by necessity 
a crude oversimplification 
that reflects  the lack of resolved data at the appropriate resolution (i.e., at a few AU scale).

\subsection{Implications for planet formation}\label{implications}

The notion that the properties of some transition disks are signposts of ongoing planet formation (e.g., Najita et al. 2007;  Paper I;   
Dodson-Robinson \& Salyk   2011; Espaillat et al. 2012) 
has gained credence from the recently identified companions to T Cha and LkCa 15. 
Using the aperture masking technique on the Very Large Telescope, Hu\'elamo et al. (2011) detected
a faint object  within the inner cavity  of the T Cha disk.  The object is located at 62 mas ($\sim$7 AU) from the primary and 
has a luminosity ratio of  5.1 mag in the L$'$-band (3.8 $\mu$m).  Based on  the upper limits from  similar K$_{s}$-band (2.2 $\mu$m) observations,
the authors derived  a K$_{S}$-L$'$ color $>$ 1.25 for the companion, suggesting the object must be surrounded by dust. 
Also using the aperture masking technique, but in the Keck telescope, Kraus \& Ireland (2012),  identified a similar object inside the 
inner hole of LkCa 15. In this case, the companion has been detected in multiple epochs and at multiple wavelengths. 
The object seems to be a point source at 2.1 $\mu$m (6.8 mag fainter than the primary), but is extended at 3.7 $\mu$m. 
Since the inclination of the  LkCa~15 disk is known from resolved submillimeter observations (Andrews et al. 2011b),  
a deprojected separation of $\sim$15 AU from the primary can be derived for the companion, assuming it  is coplanar with the 
disk. Kraus \& Ireland interpreted their observations as a young planet 
surrounded by warm dust.

RX~\,J1633.9-2442  shares an intriguing property with T Cha and LkCa 15:  a very low accretion rate  for a given disk 
mass.  
LkCa 15 has a disk mass of $\sim$55 M$_{JUP}$ (Andrews et al. 2011b) and an accretion rate of $\sim$10$^{-9}$ M$_{\odot}$/yr (Hartmann et
al.  1998). Similarly, T Cha has a disk mass of  $\sim$17 M$_{JUP}$  (Olofsson et al. 2011) and  it seems to be accreting only very 
weakly and sporadically onto the star (Alcala et al. 1993; Schisano et al. 2009).
In the absence of a planet, the mass accretion rate  onto the star should be roughly proportional to the mass of the disk (Najita et al. 2007);  
however, a planet massive enough to open a gap in the disk is expected to divert most of the material accreting from the outer disk onto itself. 
 As a result, in the presence of a Jupiter mass planet, the accretion \emph{onto the star} is reduced by a factor of $\sim$10 with 
respect to the mass transport  across the outer disk (Lubow $\&$ D$'$Angelo, 2006). 
In \S~\ref{modeling} we derived a disk mass of  $\sim$15 M$_{JUP}$ for RX~\,J1633.9-2442.  
Despite this large disk mass,  we estimate an accretion rate of $\lesssim$ 10$^{-10.0}$ M$_{\odot}$yr$^{-1}$ based on the velocity dispersion 
of its H${\alpha}$ line (see \S~\ref{magellan}).  
For comparison, the median disk mass and accretion rate for classical T Tauri stars in Ophiuchus are 5 M$_{JUP}$ (Andrews $\&$ Williams, 2005) 
and 10$^{-8.5}$ M$_{\odot}$yr$^{-1}$ (Natta et al. 2006), respectively.
As discussed in the previous section, the unusually low accretion rate onto the star for the given disk mass observed 
in RX~\,J1633.9-2442 is consistent with the presence of actively accreting objects embedded within the disk.

Moderate to low accretion rates seem to be a general feature of transition 
disks. While the distributions of mass accretion rates of transition and
non-transition disks are very wide and clearly overlap, the accretion rates of 
disks with evidence  for holes and gaps tend to be a factor of 5 lower than those 
of ``full" disks"  (Espaillat et al. 2012).
All things considered, the properties of   RX~\,J1633.9-2442,  T Cha,  and LkCa 15 are best explained by  the runaway
gas accretion phase in models of giant planet formation through core accretion (Lissauer $\&$ Stevenson, 2007).
According to the  core accretion model,  gas giant planets form by first accreting a solid core, which later attracts a massive gaseous 
envelope.  Initially the gas accretion rate onto the core is much lower than the accretion rate of solids. 
As the mass of the core increases, the rate of gas accretion accelerates. 
Once the mass of gas in the core matches the mass of solids, runaway gas accretion occurs and is sustained 
for as long as there  is material in the gas feeding zone of the planet  (i.e., until the forming planet clears a gap 
in the disk). 
Due to accretion shock luminosity, the envelope accretion phase corresponds to the highest luminosity state 
a giant planet will ever have (Marley et al. 2007). This luminosity spike could  be  in fact what makes 
the detection of forming planets even possible with current instrumentation.
The duration of the runaway gas accretion phase is estimated to be of the order of 10$^{5}$ years for a 1  M$_{JUP}$
planet  and somewhat longer for a larger planet (Marley et al. 2007). This timescale is also in agreement
with the incidence of transition disks sharing the properties of RX~\,J1633.9-2442 (combining 
a sharp inner hole, a large disk mass, and a low accretion rate).
 We note that, while the occurrence rate of objects  that can be broadly defined as transition disks is of the order of 20$\%$,
 they are \emph{not} all consistent with giant planet formation.
Accreting objects with steeply rising mid-IR SEDs, such as RX~J1633.9-2442,  T Cha, and LkCa 15 are an order of magnitude less common
(Cieza et al. 2010, 2012; Romero et al. 2012).

If our interpretation is correct,  massive accreting transition disks with rising mi-IR SEDs 
around single stars are \emph{by far} the best places for direct imaging searches of 
forming planets since they are  not only the sites of ongoing 
planet formation, but also the places where forming planets should be the brightest!  
As shown by Hu\'elamo et al. (2011) and Kraus $\&$ Ireland (2012),  non-redundant 
aperture masking is the most promising technique for such searches as it 
delivers the highest contrast ratio at the diffraction limit of the telescope. 

In addition to core accretion, gravitational instability (GI) has also been proposed
as a formation mechanism for giant planets (Boss 1997; Durisen et al. 2007). Nevertheless, 
GI seems to be less relevant to the transition disks discussed herein for several reasons. 
First, GI planets are expected to form  at evolutionary stages much earlier (age $<$ 1 Myr) than 
those of transition disks, when the disk is still extremely massive (M$_{disk}$/M$_{star}$ $\gtrsim$ 0.1)
and  deeply embedded within an extended envelope, while 
the age distribution of transition disks likely to harbor forming planets 
favors a $\gtrsim$ 2-3 Myr formation timescale (Cieza et al. 2012). 
Second,  GI should operate mostly at large radii where the cooling times are shorter 
than the local  orbital periods,  a condition needed for fragmentation (Gammie 2001, 
Rafikov 2007).  
GI is believed to be much less effective at disk radii 
$\lesssim$ 40 AU (Boley $\&$ Durinsen, 2008) or even $\lesssim$ 100 AU (Boley 2009), which makes 
GI  less  consistent  with the inner hole sizes of most transition disks. 
Finally, since the timescale for the formation of planets through GI is  
$\sim$10$^3$ yr, it is statistically unlikely that the formation event itself 
 would be observed in a nearby molecular cloud with an age of few million years 
and hundreds,  not thousands, of young stellar objects. 
Since GI models do  not run long enough to predict the long term 
evolution of the disk after the formation of the planet, it could be argued that  a GI planet 
could remain embedded in a massive disk for a relative long period of time after
it has formed. 
However, as discussed above,  the high disk masses and low accretion rates of 
RX~\,J1633.9-2442, T\,Cha and LkCa 15, as well as the properties of the T\,Cha, 
and LkCa 15 companions, suggest that the putative planets are accreting most of their mass at the
current epoch (i.e.,  we are watching them form!).  
Even though some planets might form through GI, 
the properties and incidence of  the transition objects mentioned here are in much better agreement 
with planet formation through core accretion.

\section{Summary and Future directions}

We present continuum high resolution SMA observations of the transition disk  RX\,J1633.9-2442
and simultaneously model the entire optical to millimeter 
wavelength SED and SMA visibilities in order to constrain the structure of its disk. 
The submillimeter image reveals that the disk is highly inclined ($i \sim$50 deg) 
and has an inner cavity  $\sim$25 AU in radius. This cavity is not empty as 
some warm dust is needed to explain the excess emission observed at 
12, 22, and 24 $\mu$m. 
The mid-IR excess can be reproduced with either a narrow, optically
thick ring  at $\sim$10 AU or an optically thin region extending 
from $\sim$7 to 25 AU. 
The lack of near-IR excess emission indicates 
that the inner disk (r $\lesssim$ 5 AU) is mostly  depleted of IR-emitting grains. 
Since RX~J1633.9-2442 is a single star,  the properties of the disk 
(a complex structure, a relative massive outer disk, and weak accretion) 
favors dynamical clearing by multiple planets as the hole formation 
mechanism. 
This conclusion can be extended to objects with similar 
properties and disk structures, such as the three transition 
objects mentioned in Section~\ref{mod-results}: DM Tau,  
RX~J1615.3-3255, and RX~J1604.3-2130.

The  filamentary structures predicted by hydrodynamical models of multiple planets
embedded within a disk can reconcile the accretion rate and SED of 
RX~J1633.9-2442 as they can transport significant amounts of material 
to the inner disk without overproducing the observed IR excess.
The properties and occurrence  rate of objects such as 
RX~J1633.9-2442, T Cha, and LkCa 15 (and those of the companions recently 
identified to these latter objects)  are in good agreement  with the runaway gas accretion phase of the core 
accretion model, when giant planets  gain their envelopes and suddenly become massive enough to 
dynamically clear a gap in the disk.

If the inner holes of   RX~\,J1633.9-2442,   DM Tau,  RX~J1615.3-3255,  RX~J1604.3-2130,
T Cha,  and LkCa 15 are in fact  due to ongoing giant planet formation through core accretion, 
these types of systems would represent ideal laboratories to study this complex process in 
detail and place much needed observational constraints. For instance, the location of the LkCa 15 
companion and the age of the system  would  already imply that core accretion can actually
form giant planets at $\sim$15 AU within $\sim$3 Myr, which is a difficult challenge for 
current models (Dodson-Robinson $\&$ Bodenheimer, 2010). 
Similarly, the sizes of the inner holes in the disks of 
RX~\,J1633.9-2442,  RX J1615.3-3255,  and  RX J1604.3-2130 
also suggest the presence of young giant planets 
at $\gtrsim$ 20 AU orbital separations.

In the near future, the unprecedented sensitivity and resolution of the
Atacama Large Millimeter/submillimeter Array (ALMA) will revolutionize 
the fields of disk evolution and planet formation. 
Most studies of nearby circumstellar disks  will soon be based on high resolution images 
of both thermal emission and molecular gas tracers rather than on SED observations. 
ALMA will provide new insights on the structure of disks and their dynamics
and teach us about turbulence, grain growth and dust settling, and the evolution 
of  the dust to gas mass ratio, the understanding of all of which are key to planet
formation theory. 
Detailed ALMA studies of disks hosting forming planets 
is the most direct  and promising approach to learn about the planet formation
process and the conditions in which planets form.

\acknowledgments
Support for this work  was provided by NASA through 
the \emph{Sagan} Fellowship Program  under an
award from the California Institute of Technology. 
G.S.M. and J.P.W. acknowledge NASA/JPL and NSF for funding 
support through grants RSA-1369686 and AST08-08144 respectively.
A.L.K. was supported  by NASA through the \emph{Hubble} Fellowship
program. 
M.R.S and F.D.M acknowledge support from Millenium Science Initiative, 
Chilean Ministry of Economy, Nucleus P10-022-F. 

\newpage

\begin{deluxetable}{crrlrr}
\footnotesize
\tablecaption{ RX\,J1633.9-2442 photometry data}
\tablehead{\colhead{Wavelength}&\colhead{Flux}& \colhead{Flux}& \colhead{Error$^a$}&\colhead{Telescope}&\colhead{Reference$^b$}\\
                    \colhead{($\mu$m)}&\colhead{(mJy)}&\colhead{(mag)} &\colhead{(mJy)}& \colhead{}&\colhead{} }
\startdata
  0.65  &   3.53e+00  &   14.85 &    30$\%$ &   USNO-B1 &   1   \\
 1.2    &   1.04e+02    &  10.46  &  15$\%$    & 2MASS  &   2 \\
  1.6  &  1.85e+02   &      9.36   &   15$\%$       &   2MASS & 2 \\
  2.2  &  2.02e+02   &      8.80   &    15$\%$   &  2MASS & 2 \\
  3.6   &  9.67e+01  &     8.66    &   10$\%$       & \emph{Spitzer} & 3 \\
  4.5  &   7.10e+01  &     8.51    &    10$\%$    & \emph{Spitzer} & 3 \\
  5.8  &   5.12e+01  &      8.37    &   10$\%$     & \emph{Spitzer} & 3\\
  8.0  &   3.28e+01 &       8.23    &   10$\%$      & \emph{Spitzer} & 3 \\
  12   &   2.41e+01  &      7.80      &  10$\%$                &    WISE & 4 \\
  22   &   2.54e+02  &       3.79      &  10$\%$                &   WISE  & 4  \\
  24   &   2.28e+02   &      3.74       & 10$\%$  & \emph{Spitzer}  & 3 \\
  70   &    7.13e+02   &   \nodata   & 15$\%$  &  \emph{Spitzer}   &  3\\
  850 &  2.10e+02    &   \nodata   &15$\%$  &           JCMT           & 5 \\
1300 &   8.18e+01   &  \nodata   &15$\%$  &             SMA           &  3 \\ 
\enddata
\tablecomments{
$^a$the optical and near-IR uncertainties are dominated by the extinction corrections.
$^b$References:  (1)  Monet et al. 2003;  (2) Skrutskie et al.  (2006);  (3)  Cieza et al. (2010); 
(4) Cutri et al. 2011;
(5) Nutter et al. (2006).}
\end{deluxetable}

\begin{deluxetable}{lr}
\footnotesize
\tablecaption{Stellar parameters and fixed disk parameters}
\tablehead{\colhead{Parameter}&\colhead{Adopted value}} \\
\startdata
Stellar T$_{eff}$         [K]                       & 4350     \\
log g                                                 &        4 \\
Distance        [pc]                           &   120 \\
Inclination     [deg]                         &  50 \\
Grain size distribution slope, p   &  $-$3.5     \\
 a$_{min}$  [$\mu$m ] &  0.005     \\
 a$_{max}$ [$\mu$m] &  3900        \\
\enddata\label{fix_table}
\end{deluxetable}

\begin{deluxetable}{lccc}
\footnotesize
\tablecaption{Fitted disk parameters and $\chi^2$ statistics}
\tablehead{\colhead{Parameter}&\colhead{Empty cavity}&\colhead{2-component disk}&\colhead{Ring within cavity}} \\
\startdata

$R_{C}$  [AU]                                &   39.7 $\pm$ 5.6            &  38.2 $\pm$ 5.3                 &  42.7 $\pm$ 3.7  \\
$R_{cav}$                                       &   22.7 $\pm$ 1.6            &  25.6  $\pm$  1.4                & 27.3  $\pm$ 2.1 \\  
log($\Sigma_C$)[g/cm$^2$]       &  $-$0.23  $\pm 0.40$   &  $-$0.31 $\pm$ 0.31     &   $-$0.39  $\pm$ 0.24        \\
$\gamma$                                       &   -0.49 $\pm$ 0.24         &    $-$0.30 $\pm$ 0.41          &  $-$0.81$\pm$ 0.16              \\
$h_C$                                              &    0.014 $\pm$ 0.003 &  0.018  $\pm$ 0.002    &    0.019 $\pm$ 0.002     \\ 
$\psi$                                                &   0.37   $\pm$ 0.04   &  0.30  $\pm$  0.09        &     0.22  $\pm$  0.14       \\ 
$R_{in}$      [AU]                             &     \nodata       &   6.9 $\pm$ 1.6       &     \nodata         \\ 
log($\delta_{cav}$)                        &  \nodata         &   $-$2.0  $\pm$ 0.4   &     \nodata         \\ 
$R_{ring,in}$      [AU]                     & \nodata           &   \nodata                    &  10.0  $\pm$ 2.3   \\
$R_{ring,width}$    [AU]                 & \nodata           &   \nodata                    &   2.2 $\pm$  0.15  \\
$M_{disk} [$M$_{JUP}$]$^a$        &   19                &     13                        &    14     \\                                      
$\chi^2$ Visibility (range)$^b$                        &  14--38      &   12--27  &     26--50    \\
$\chi^2$ SED       (range)                                  &   89--109   &   24--33   &    27--33    \\
$\chi^2$ Visibility (adopted)$^c$                     &     36    &    21  &   28     \\
$\chi^2$ SED       (adopted)                               &   103   &    24  &   31     \\
$\chi^2$ Total      (adopted)                               &   139   &    45  &   59     \\
\enddata
%\footnotetext[1]
\tablecomments{$^a$
$M_{disk}$ is \emph{not} a free parameter. It is obtained by integrating the 
 surface density profile over radius and assuming a gas to dust mass ratio of 100.   
%\footnotemark[2]
$^b$The $\chi^2$ range corresponds to the best-fit models of the five runs of the search algorithm 
for each disk parameterization. $^{c}$ The ``adopted'' $\chi^2$ value is that of the model that gives the
best \emph{global} fit to the data (i.e., the minimum value for  $\chi^2$ Total =  $\chi^2$ Visibility + $\chi^2$ SED).} 
\end{deluxetable}

\begin{figure}
\figurenum{1}
%trim option's parameter order: left bottom right top
 \includegraphics[width=15cm, trim =0mm 0mm 2mm 2mm, clip]{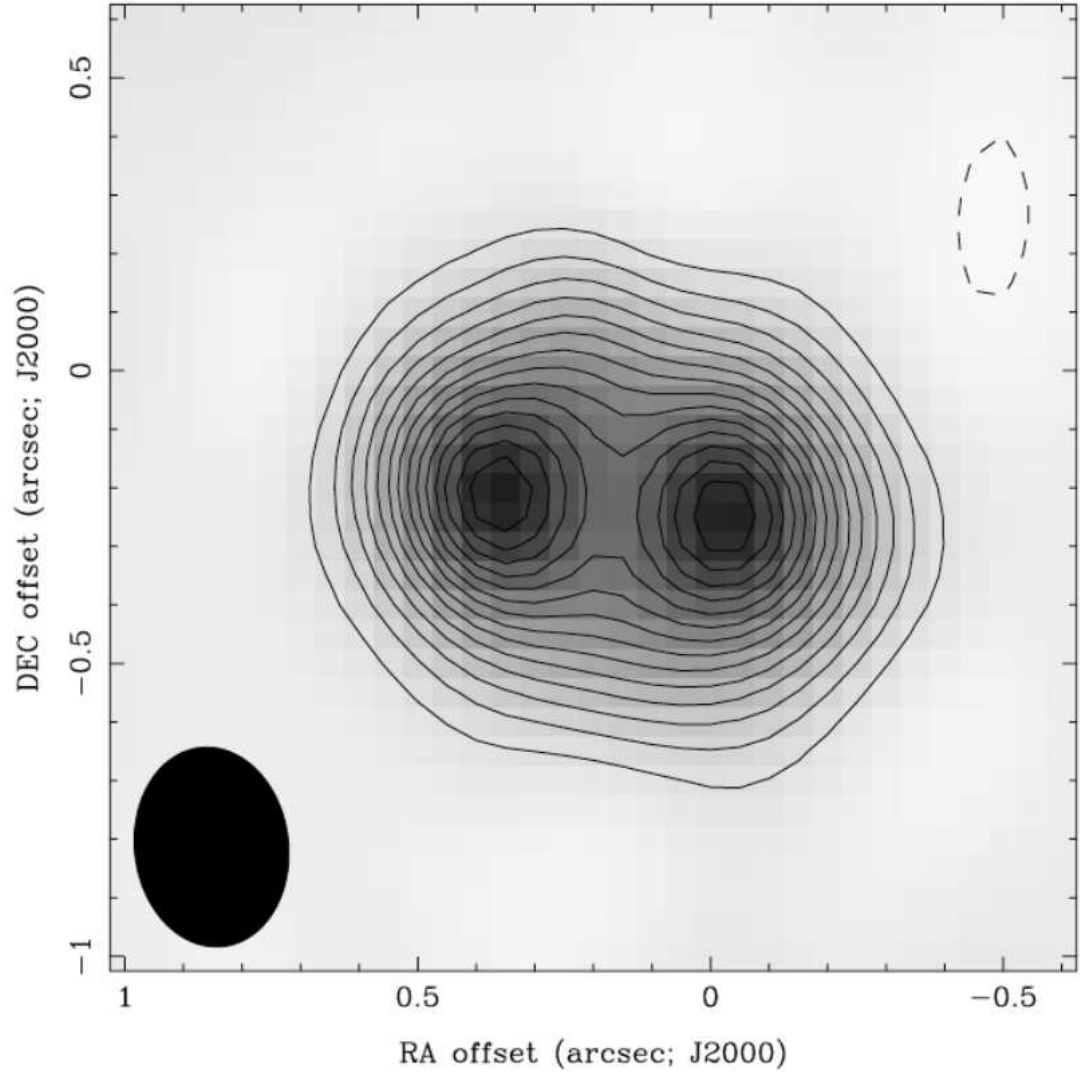}
\caption{The 340 GHz dust continuum image of  RX J1633.9-2442, combining two full very extended
SMA tracks.  The synthesized beam, 0.26$''$$\times$0.31$''$ in size (i.e.,  31 AU 
$\times$ 37 AU), is shown at the bottom left. 
The contours shown are in steps of 3 mJy/beam and start at 3 mJy/beam ($\sim$3-$\sigma$). 
The image shows two peaks because the inner hole
has been resolved in one direction (East-West),  but not in the other, suggesting 
an inclined disk. 
}
\end{figure}

\begin{figure}
\figurenum{2}
%trim option's parameter order: left bottom right top
\includegraphics[width=17cm, trim =0mm 1mm 0mm 0mm, clip]{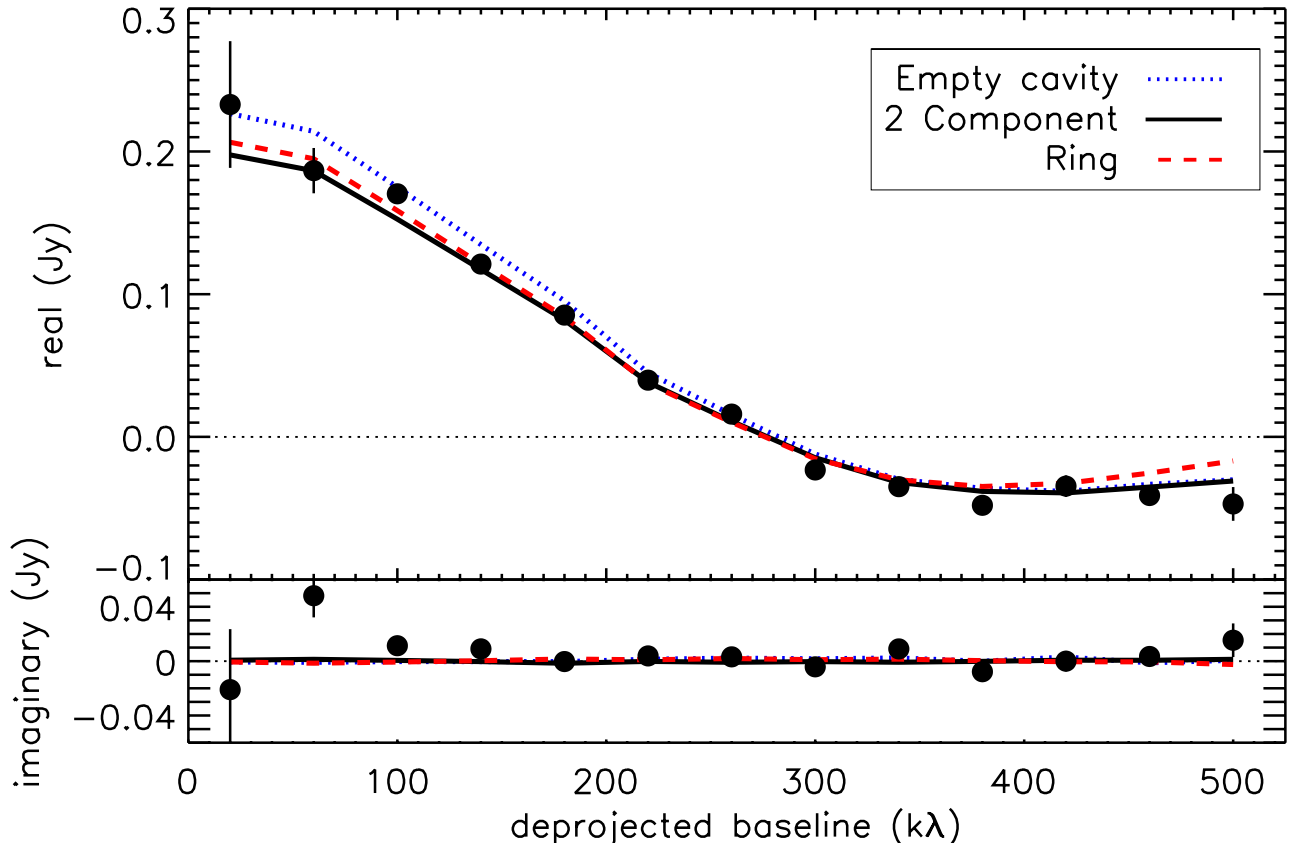}
\caption{
Deprojected visibility profile  (visibility flux vs deprojected baseline 
length). A null is seen at a baseline length of  $\sim$270-280 k$\lambda$  (240-250 m), indicating 
a sudden density change  at  $\sim$25 AU.
The deprojected visibilities of the 3 models discussed in Section~\ref{paramet} are overlaid.
} 
\end{figure}

\clearpage

\begin{figure}
\figurenum{3}
%trim option's parameter order: left bottom right top
 \includegraphics[width=15cm, trim =0mm 0mm 0mm 0mm, clip]{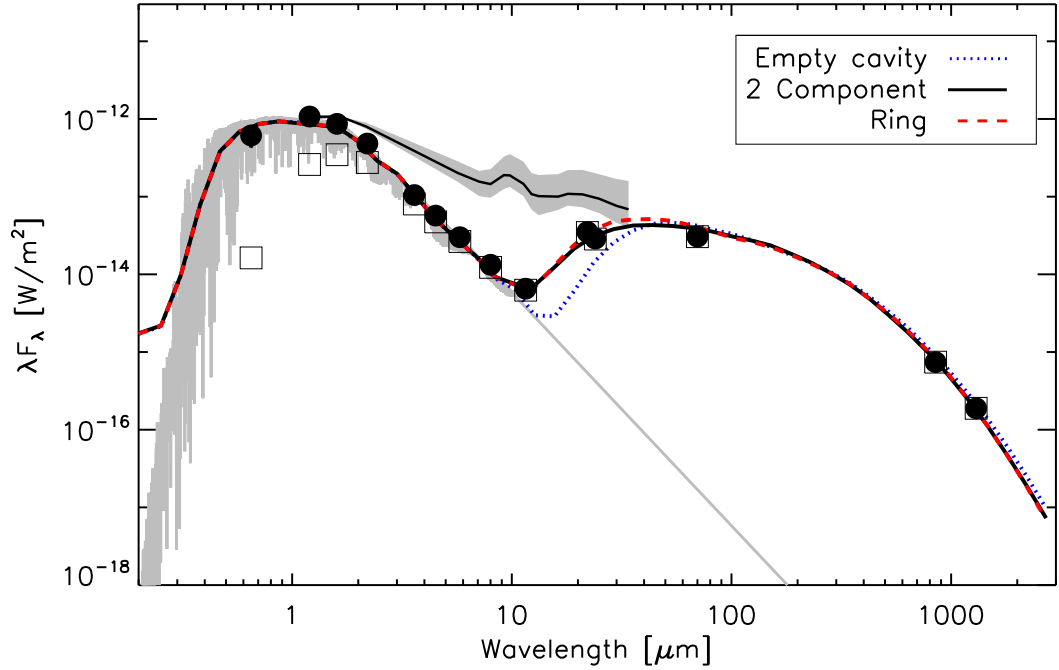}
\caption{
The optical to millimeter wavelength SED of RXJ1633.9-2442. 
 Open squares correspond to the observed fluxes before being corrected for extinction. 
 The  filled circles are the extinction-corrected  fluxes. 
The solid gray line represent the stellar photosphere (Kurucz model).
The solid gray region encompasses the median (black line) 
and 50$\%$ of the IR SEDs of  K5--M2 CTTSs  (Furlan
et al.  2006). The deficit of near- and mid-IR excesses in our target is quite obvious. 
The SEDs of the 3 models discussed in Section~\ref{paramet} are overlaid. 
}
\end{figure}

\begin{figure}
\figurenum{4}
\includegraphics[width=8cm, trim =20mm 0mm 30mm 0mm, clip]{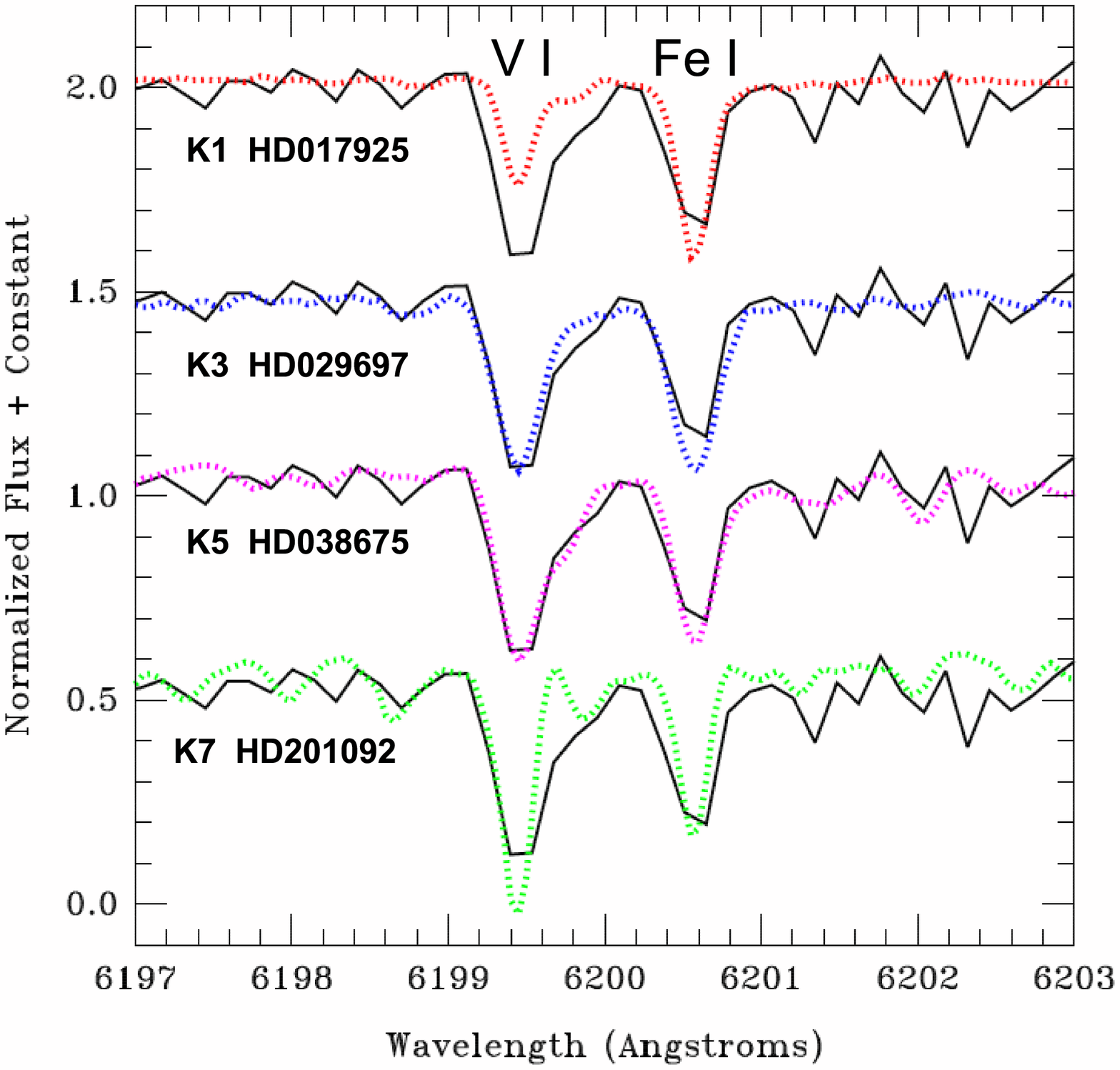}
\includegraphics[width=10cm, trim =0mm 0mm 0mm 0mm, clip]{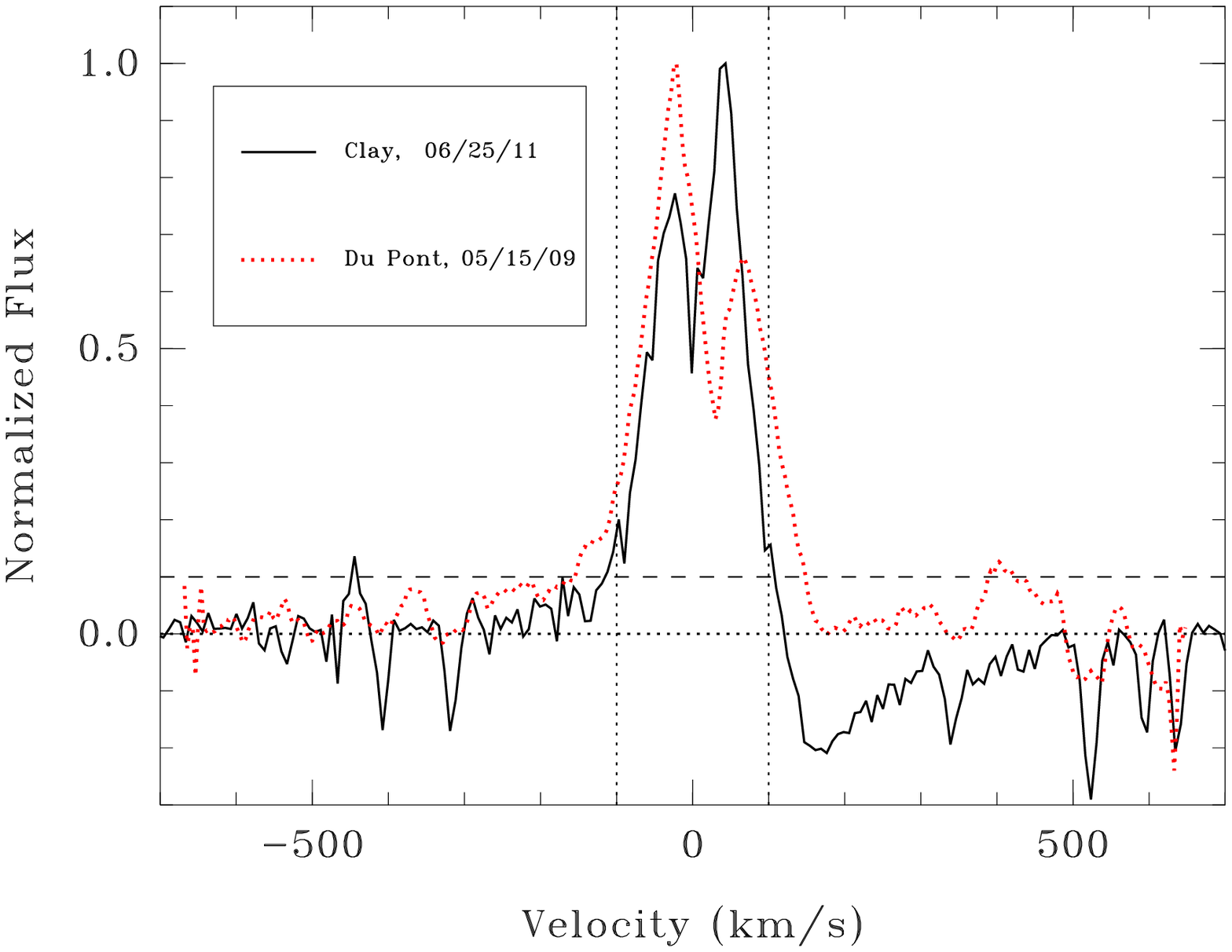}
\caption{\textbf{Left:} The observed spectrum of RX J1633.9-2442 around 6200 \AA  (solid line) 
showing the V I and Fe I  absorption lines, the ratio of  which are highly sensitive to  the effective temperature of
 K-type stars. Five spectral standards (doted lines) are shown for comparison. The K5 star is 
 the best match. \textbf{Right:}  The continuum subtracted  H$\alpha$ velocity profile of RX J1633.9-2442 two years apart. 
 The dashed line
 indicates 10$\%$ peak intensity level, where the velocity width is measured. 
 The H$\alpha$ line is asymmetric, variable,  and $>$200 km/s wide, indicating accretion onto the 
star, although at very low rates (10$^{-10.0}$ to 10$^{-10.6}$
 M$_{\odot}$yr$^{-1}$) } 
\end{figure}

\begin{figure}
\figurenum{5}
%trim option's parameter order: left bottom right top
 \includegraphics[width=17cm, trim =0mm 0mm 0mm 0mm, clip]{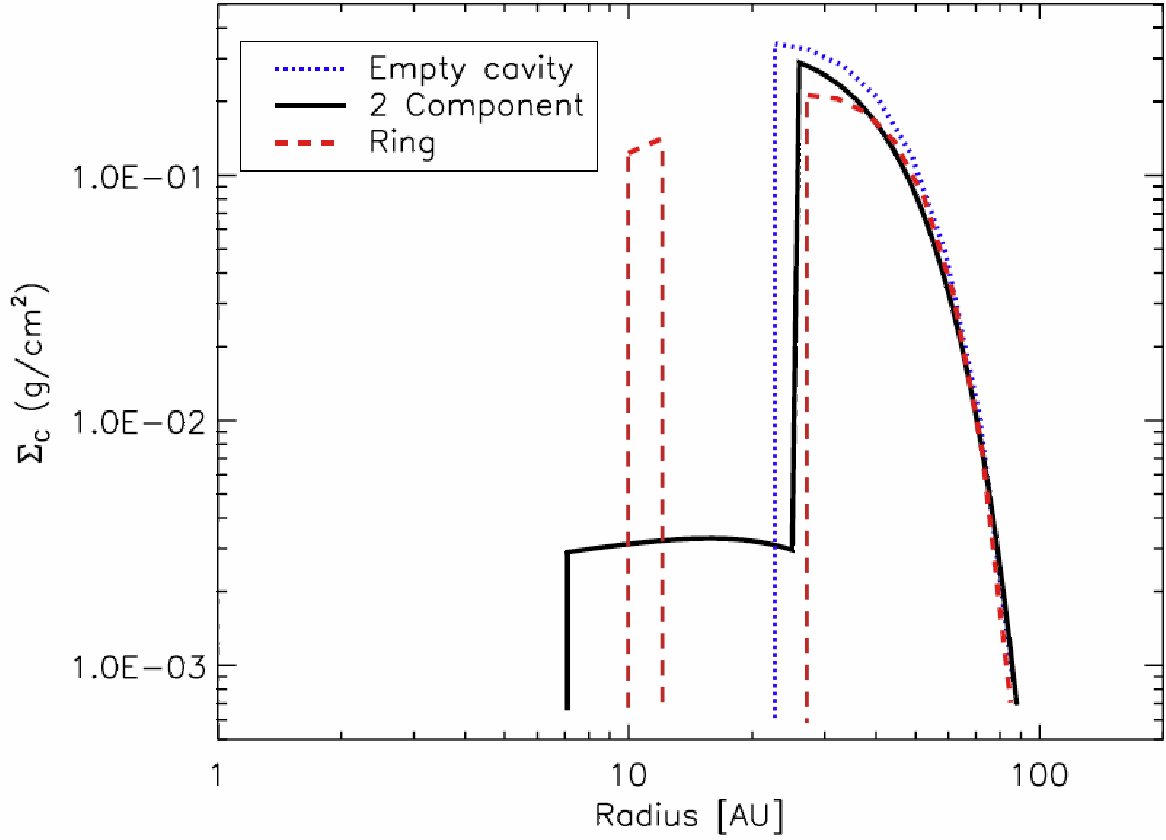}
\caption{
The surface density function of the 3 models discussed in Section~\ref{paramet}.
The lack of IR excess shortward of $\sim$10 $\mu$m indicates that the inner $\sim$7-10 AU of the 
disk is depleted of small dust grains.  The submillimeter image shows a sharp drop of the surface 
density of the disk at $\sim$25 AU; however, the 12, 22, and 24 $\mu$m fluxes imply
that the submillimeter hole is not completely empty and some warm dust must be present at distances 
of the order of 10 AU. The ``empty cavity"  disk model can thus be ruled out. 
The surface density of the disk is not well constrained beyond  a  radius of $\sim$60 AU 
as the submillimeter emission from the outer disk  quickly falls below the noise of the SMA image.  
}
\label{sigma}
\end{figure}

\end{document}